\documentclass[aps,showpacs]{revtex4}
\newtheorem{thm}{Theorem}

\newcommand{\B}{\mathcal{B}}

\newcommand{\rank}{\mathrm{rank}}
\newcommand{\Tr}{\mathrm{Tr}}

\newcommand{\ben}{\begin{enumerate}}
\newcommand{\een}{\end{enumerate}}
\newcommand{\bit}{\begin{itemize}}
\newcommand{\eit}{\end{itemize}}
\newcommand{\be}{\begin{equation}}
\newcommand{\ee}{\end{equation}}
\newcommand{\bdm}{\begin{displaymath}}
\newcommand{\edm}{\end{displaymath}}
\newcommand{\bea}{\begin{eqnarray}}
\newcommand{\eea}{\end{eqnarray}}
\newcommand{\f}[1]{\fbox}

\newcommand{\realnos}{\mbox{{\bf R}}}

\newcommand{\integers}{\mbox{{\bf Z}}}
\newcommand{\naturalnos}{\mbox{{\bf N}}}

\newcommand{\dfrac}[2]{\displaystyle{\frac{#1}{#2}}}

\begin{document}

\title{Effects of Noise on Entropy Evolution}
\author{Michael C. Mackey}
\email{mackey@cnd.mcgill.ca} \affiliation{Departments of Physiology, Physics \& Mathematics and Centre for
Nonlinear Dynamics, McGill University, 3655 Promenade Sir William Osler, Montreal, QC, CANADA, H3G 1Y6}
\author{Marta Tyran-Kami\'nska}
\email{mtyran@us.edu.pl}
\thanks{Corresponding author}
\affiliation{Institute of Mathematics, Silesian University, ul. Bankowa 14, 40-007 Katowice, POLAND}
\date{\today}

\begin{abstract}
We study the  convergence properties of the conditional (Kullback-Leibler)
entropy in stochastic systems. We have proved very general results showing that
asymptotic stability is a necessary and sufficient condition for the monotone
convergence of the conditional  entropy to its maximal value of zero.
Additionally we  have made specific calculations of the rate of convergence of
this entropy to  zero in a one-dimensional situations, illustrated by
Ornstein-Uhlenbeck and Rayleigh processes, higher dimensional situations, and
a two dimensional Ornstein-Uhlenbeck process  with a stochastically perturbed
harmonic oscillator and colored noise as examples. We also apply our general
results to the problem of conditional entropy convergence in the presence of
dichotomous noise.  In both the single dimensional and multidimensional cases
we are to show that the convergence of the conditional entropy to zero is
monotone and at least exponential. In the specific cases of the
Ornstein-Uhlenbeck and Rayleigh processes as well as the stochastically
perturbed harmonic oscillator and colored noise examples, we have the rather
surprising result that the rate of convergence of the entropy to zero is
independent of the noise amplitude $\sigma$ as long as $\sigma > 0$.
\end{abstract}
 \maketitle

\section{Introduction}\label{s:intro}

This paper examines the role of noise in the evolution of the conditional (or
Kullback-Leibler or relative) entropy to a maximum.  We were led to examine
this problem because it is known that in invertible systems (e.g. measure
preserving systems of differential equations or invertible maps) the
conditional entropy is fixed at the value with which the system is prepared
\citep{andrey85,steeb79,mcmtdbk,mcmitaly}, but that the addition of noise can
reverse this invertibility property and lead to an evolution of the conditional
entropy to a maximum value of zero. Here, we make both general and concrete
specific calculations to examine the entropy convergence. We carry this out by
using convergence properties of the Fokker-Planck equation  using `entropy
methods' \citep{evans04}, which  have been known for some time to be useful for
problems involving questions related to convergence of solutions in partial
differential equations \citep{loskot91,abbond99,arnold01,qian01,qian02}. Their
utility can be traced, in some instances, to the fact that entropy may serve as
a Liapunov functional \citep{markowich00}.




The outline of this paper is as follows.   Section \ref{s:asre} introduces the dynamic concept of asymptotic
stability and the notion of conditional entropy. (Asymptotic stability is a strong convergence property of
ensembles which implies mixing.  Mixing, in turn, implies ergodicity.) This is followed by two main results
connecting the convergence of the conditional entropy with asymptotic stability (Theorem \ref{t:entropyconv}),
and the existence of unique stationary densities with the convergence of the conditional entropy (Theorem
\ref{t:unique}). Section \ref{s:det} shows that asymptotic stability is a property that cannot be found in a
deterministic system of ordinary differential equations, and consequently that the conditional entropy does not
have a dynamic (time dependent) character for this type of invertible dynamics. Section \ref{s:gauss} considers
the stochastic extension where a system of ordinary equations is perturbed by Gaussian white noise (thus
becoming non-invertible) and gives some general results showing that in this stochastic case asymptotic
stability holds. Section \ref{ss:1d} starts to consider specific one dimensional examples, and we show that the
conditional entropy convergence to zero is at least exponential. We consider the Ornstein-Uhlenbeck process in
Section \ref{sss:1dou} and a Rayleigh process in Section \ref{sss:rp}.  We consider multidimensional stochastic
systems in Section \ref{ss:md}, showing that the exponential convergence of the entropy still holds. Specific
examples of these higher dimensional situations are considered within the context of a two dimensional
Ornstein-Uhlenbeck process in Section \ref{ss:mo-up} with specific examples of a stochastically perturbed
harmonic oscillator (Section \ref{sss:ho}) and colored noise (Section \ref{sss:cn}) as examples. Section
\ref{s:mdn} applies our general results to the problem of conditional entropy convergence in the presence of
dichotomous noise.  The paper concludes with a short discussion.

\section{Asymptotic stability and conditional entropy}\label{s:asre}


Let $(X,\B,\mu)$ be a $\sigma$-finite measure space. Let $\{P^t\}_{t\ge 0}$  be a semigroup of Markov operators
on $L^1(X,\mu)$, {\it i.e.} $P^tf\ge 0$ for $f\ge 0$, $\int P^tf(x)\, \mu(dx)=\int f(x) \, \mu(dx)$, and
$P^{t+s}f=P^t(P^sf)$. If the group property holds for $t,s \in \realnos$, then we say that $P$ is invertible,
while if it holds only for $t,s \in \realnos ^+$ we say that $P$ is non-invertible. We denote the corresponding
set of densities by $\mathcal{D}(X,\mu)$, or $\mathcal{D}$ when there will be no ambiguity, so $f \in
\mathcal{D}$ means $f \geq 0$ and $\int_X f(x) \,\mu(dx) = 1$. We call a semigroup of Markov operators $P^t$ on
$L^1(X,\mu)$ {\it asymptotically stable} if there is a density $f_*$ such that $P^tf_*=f_*$ for all $t>0$ and
for all densities $f$
$$
\lim_{t\to\infty}P^tf=f_*.
$$
The density $f_*$ is called a stationary density of $P^t$.

We define the
conditional entropy (also known as the Kullback-Leibler or relative entropy) of
two densities $f,g\in \mathcal{D}(X,\mu)$ as
    \be
    H_c(f|g)=-\int_X f(x)\log\dfrac{f(x)}{g(x)}\mu(dx). \label{d:conent}
    \ee
Our first result connects the temporal convergence properties of $H_c$ with those of $P^t$.


%

\begin{thm}\label{t:entropyconv}
Let  $P^t$ be a semigroup of Markov operators on $L^1(X,\mu)$ and $f_*$ be
a positive density. If
    \be
    \lim_{t \to \infty} H_c(P^tf|f_*) = 0
    \label{c:entconv}
    \ee
for a density $f$ then
    \be
    \lim_{t \to \infty} P^t f = f_*.
    \label{c:l1conv}
    \ee
Conversely,  if $P^tf_*=f_*$ for all $t\ge 0$  and Condition \ref{c:l1conv} holds for all $f$ with
$H_c(f|f_*)>-\infty$, then Condition \ref{c:entconv} holds as well.
\end{thm}

{\noindent\bf Proof.}  To prove the first part, note that for densities $f,g
\in \mathcal{D}$, we have from the Csisz\'ar-Kullback inequality
    $$
    -||f-g||_{L^1}^2 \geq 2 H_c(f|g).
    $$
In particular,
    $$
    -||P^t f-f_*||_{L^1}^2 \geq 2 H_c(P^t f|f_*).
    $$
Thus if
    $$
    \lim_{t \to \infty} H_c(P^t f|f_*) = 0,
    $$
then
    $$
    \lim_{t \to \infty} ||P^t f - f_*||_{L^1} = 0
    $$
and $P^t$ is asymptotically stable.

Proof of the converse portion is not so straightforward. Assume initially that $f/f_*$ is bounded, so  $0\le
f\le a f_*$. Since $P^t f_*=f_*$ and $P^t$ is a positive preserving operator, we have $0\le P^tf\le a f_*$.
Making use of the inequality, proved in the appendix, 
    \be
    -H_c(f|g)\le \int \left(\frac{f}{g}-1\right)^2g\, \mu(dx),\label{i:foot1}
    \ee
valid for all densities $f,g$, we arrive at
    \bea
    -H_c(P^tf|g)& \le &\int \left|\frac{P^tf}{f_*}-1\right|
    |P^tf-f_*|\, \mu(dx)\nonumber \\
    &\le &\max\{1,|a-1|\}||P^tf-f_*||_1.\nonumber
    \eea
Consequently, $H_c(P^tf|f_*)\to 0$ as $t\to\infty$ for every density $f$ such that $f/f_*$ is bounded.

Now assume that $H_c(f|f_*)>-\infty$ and write $f$ in the form $f=g_a+f_a$ where $g_a=f-f_a$ and
    $$
    f_a(x)=\left\{
    \begin{array}{ll} 0, & f(x)>a f_*(x),\\
    f(x), &0\le f(x)\le a f_*(x).
    \end{array}\right.
    $$
We have $1=||f||_1=||g_a||_1+||f_a||_1$ and $||g_a||_1=\int_{\{f>a f_*\}}f\mu(dx)\to 0$ as $a\to\infty$. Write
$P^tf$ in the form
    \be
    P^tf=||g_a||_1 P^t\left(\dfrac{g_a}{||g_a||_1}\right)+||f_a||_1
    P^t\left(\dfrac{f_a}{||f_a||_1}\right). \label{e:deco}
    \ee Since $\dfrac{f_a}{f_*||f_a||_1}$ is bounded, we have
$H_c\left (\dfrac{f_a}{||f_a||_1}|f_*\right )>-\infty$ and
    $$
    \lim_{t\to\infty}H_c\left(P^t\left(\dfrac{f_a}{||f_a||_1}\right)|f_*\right)=0
    $$
 by the first part of our proof.  Observe that
    $$
    H_c\left(\dfrac{g_a}{||g_a||_1}|f_*\right)=\int_{\{f>a f_*\}}f
    \log\frac{f}{f_*}\mu(dx),
    $$
which, because of the condition $H_c(f|f_*)>-\infty$,  is convergent to $0$ as $a\to\infty$. Furthermore
$H_c(P^tf|f_*)\ge H_c(f|f_*)$ for any density $f$ and the function $f\mapsto H_c(f|f_*)$ is concave. Combining
these results and equality \ref{e:deco} we obtain
   \begin{widetext} \begin{eqnarray*}
    H_c(P^tf|f_*)& \ge & ||g_a||_1
    H_c\left(P^t\left(\dfrac{g_a}{||g_a||_1}\right)|f_*\right)+||f_a||_1
    H_c\left(P^t\left(\dfrac{f_a}{||f_a||_1}\right)|f_*\right)\\
    & \ge & ||g_a||_1
    H_c\left(\dfrac{g_a}{||g_a||_1}|f_*\right)+||f_a||_1
    H_c\left(P^t\left(\dfrac{f_a}{||f_a||_1}\right)|f_*\right).
    \end{eqnarray*}
    \end{widetext}
Letting $t \to \infty$, followed by taking the limit $a\to \infty$ completes the proof of the {\it if} part, and
of the theorem.



The semigroup $P^t$ on $L^1(X,\mu)$ is called {\it partially
integral} if there exists a measurable function $q : X \times X \to
[0,\infty)$ and $t_0>0$ such that
   $$
   P^{t_0}f(x) \ge \int_X q(x, y)f(y)\,\mu(dy)
    $$
for every density $f$ and
$$
\int_X \int_X q(x, y)\,\mu(dy)\,\mu(dx)>0.
$$


Our next result draws a connection between the existence of a unique stationary density $f_*$ of $P^t$ and the
convergence of the conditional entropy $H_c$.
\begin{thm}\label{t:unique}
Let $P^t$ be a partially integral semigroup of Markov operators. If there is a unique stationary density $f_*$
for $P^t$ and $f_*>0$, then
$$\lim_{t \to \infty} H_c(P^tf_0|f_*) = 0$$  for all $f_0$ with
$H_c(f_0|f_*)>-\infty$.
\end{thm}
{\noindent\bf Proof.} From \citet[Theorem 2]{pichorrudnicki00} the semigroup $P^t$ is asymptotically stable if
there is a unique stationary density $f_*$ for $P^t$ and $f_*>0$. Thus the result follows from Theorem
\ref{t:entropyconv}.

\setcounter{equation}{0}
\section{Conditional entropy in invertible deterministic systems}
\label{s:det} 

In this section we briefly consider the behaviour of the 
conditional entropy in situations where the dynamics are invertible in the
sense that they can be run forward or backward in time without ambiguity. To
make this clearer, consider a phase space $X$ and a dynamics $S_t:X \to X$. For
every initial point $x^0$, the sequence of successive points $S_t(x^0)$,
considered as a function of time $t$, is called a {\bf trajectory}. In the
phase space $X$, if the trajectory $S_t(x^0)$ is nonintersecting with itself,
or intersecting but periodic, then at any given final time $t_f$ such that $x^f
= S_{t_f}(x^0)$ we could change the sign of time by replacing $t$ by $-t$, and
run the trajectory backward using $x^f$ as a new initial point in $X$. Then our
new trajectory $S_{-t}(x^f)$ would arrive precisely back at $x^0$ after a time
$t_f$ had elapsed: $x^0 = S_{-t_f}(x^f)$.  Thus in this case we have a dynamics
that may be reversed in time {\it completely unambiguously}.  

We formalize this by introducing the concept of a {\bf dynamical system} $\lbrace S_t \rbrace _{t \in
\realnos}$, which is simply any group of transformations $S_t:X \rightarrow X$ having the two properties: 1.
$S_0 (x) = x$; and 2. $S_t(S_{t'}(x)) = S_{t+t'}(x)$ for $t,t'\in \realnos $ or $\integers$. Since, from the
definition, for any $t \in  \realnos$, we have $S_t(S_{-t}(x)) = x = S_{-t}(S_t(x))$, it is clear that dynamical
systems are invertible in the sense discussed above since they may be run either forward or backward in time.
Systems of ordinary differential equations are examples of dynamical systems as are invertible maps.

Our first result is very general, and shows that the conditional entropy of any
invertible system is constant and uniquely determined by the method of system
preparation.  This is formalized in

\begin{thm}
\label{thm-invert}
If $P^t $ is an invertible Markov operator and has a   stationary density
$f_*$, then the conditional entropy is constant  and equal to the value
determined by $f_*$ and the choice of the initial density $f_0$ for all time
$t$.  That is,
\begin{equation}
H_c (P^tf_0 |f_*) \equiv H_c(f_0|f_*) \label{e-invertent}
\end{equation}
for all $t$.
\end{thm}
\noindent \textbf{Proof.}  Since $P $ is invertible, by Voigt's theorem \footnote {Voigt's theorem
\citep{voigt81} says that if $P$ is a Markov operator, then $ H_c(P f|P g) \geq H_c(f|g) $ for all $f,g \in
\mathcal{D}$.} with $g = f_*$ it follows that
\begin{eqnarray*}
  H_c(P^{t+t'}f_0|f_*) &=& H_c(P^{t'}P^tf_0|f_*) \\
   &\geq & H_c(P^tf_0|f_*) \geq H_c(f_0|f_*)
\end{eqnarray*}
for all times $t$ and $t'$.  Pick $t' = -t $ so
$$
          H_c(f_0|f_*) \geq H_c(P^tf_0|f_*) \geq H_c(f_0|f_*)
$$
and therefore
$$
H_c(P^tf_0|f_*) = H_c(f_0|f_*)
$$
for all $t$, which finishes the proof.

In the case where we are considering a deterministic dynamics $S^t: {\cal X} \to {\cal X}$ where ${\cal X}
\subset X$, then the corresponding Markov operator is also known as the Frobenius Perron operator
\citep{almcmbk94}, and is given explicitly by
    \be
    P^tf_0(x) = f_0(S^{-t}(x)) | J^{-t}(x)|
    \ee
where $J^{-t}(x)$ denotes the Jacobian of $S^{-t}(x)$. A simple calculation shows
    \bea
    H_c(P^tf_0|f_*) &=& -\int_ {\cal X} P^t f_0(x)\log \left [ \dfrac{P^t f_0(x)}{f_*(x)}\right ]\, dx \nonumber \\
    &=& -\int_ {\cal X} f_0(S^{-t}(x))|J^{-t}(x)|\log \left [\dfrac{ f_0(S^{-t}(x)) }{f_*(S^{-t}(x))}\right ]\, dx \nonumber \\
    &=& -\int_ {\cal X} f_0(y)\log\left [ \dfrac{ f_0( y) } {f_*(y)}\right ] \, dy \nonumber \\
    &\equiv&  H_c(f_0|f_*) \nonumber
    \eea
as expected from Theorem \ref{thm-invert}.

 More specifically, if the dynamics corresponding to our invertible Markov operator are described by the
system of ordinary differential equations
    \be
    \dfrac{dx_i}{dt} = F_i(x) \qquad        i = 1,\ldots ,d
    \label{ode}
    \ee
operating in a region of $\realnos^d$ with initial conditions $x_i(0) = x_{i,0}$, then \citep{almcmbk94} the
evolution of $f(t,x) \equiv P^tf_0(x)$ is  governed by the generalized Liouville  equation
\begin{equation}
\frac {\partial f}{\partial t} = -\sum_i \frac {\partial (fF_i)}{\partial x_i}. \label{e-leqn}
\end{equation}
The corresponding stationary density  $f_*$ is given by the solution of
\begin{equation}
\sum_i \frac {\partial (f_* F_i)}{\partial x_i} = 0. \label{e-liouss}
\end{equation}
Note that the uniform density  $f_* \equiv 1$, meaning that the flow defined by Eq. \ref{ode} preserves the
Lebesque measure,  is a stationary density of Eq. \ref{e-leqn} if and only if
\begin{equation}
\sum_i \frac {\partial F_i}{\partial x_i} = 0.
\end{equation}

In particular, for the system of ordinary differential equations (\ref{ode}) whose density evolves according to
the Liouville equation (\ref{e-leqn}) we can assert that the conditional entropy of the density $P^tf_0$ with
respect to the stationary density $f_*$ will be constant for all time and will have the value determined by the
initial density $f_0$ with which the system is prepared. This result can also be proved directly by noting that
from the definition of the conditional entropy we may write
$$
H_c(f|f_*) = -  \int_{\realnos ^d}  f(x) \left[ \log \left( \frac{f}{f_*} \right) + \frac{f_*}{f} - 1 \right] \,
dx
$$
when the stationary density is $f_*$.  Differentiating with respect to time gives
$$
\frac {dH_c}{dt}   = - \int_{\realnos ^d}  \frac {df}{dt} \log \left[ \frac {f}{f_*} \right] \,dx
$$
or, after substituting from (\ref{e-leqn}) for $(\partial   f/\partial   t)$, and integrating by parts under the
assumption that $f$ has compact support,
$$
\frac {dH_c}{dt} = \int_{\realnos ^d}  \frac{f}{f_*} \sum_i \frac {\partial (f_*F_i)}{ \partial x_i} \,dx.
$$
However, since $f_*$ is a stationary density of $P^t$, it is clear from (\ref{e-leqn}) that
$$
\frac {dH_c}{dt} = 0,
$$
and we conclude that the conditional entropy $H_c(P^tf_0|f_*)$ does not change from its initial value when the
dynamics evolve in this manner.

\setcounter{equation}{0}
\section{Effects of Gaussian noise}\label{s:gauss}

In this section, we turn our attention to the behaviour of the stochastically perturbed system
    \be
    \dfrac{dx_i}{dt} = F_i(x) + \sum_{j=1}^d \sigma_{ij}(x) \xi _j, \qquad   i = 1,\ldots ,d
    \label{stochode}
    \ee
with the initial conditions $x_i(0) = x_{i,0}$,  where   $\sigma _{ij}(x)$ is the amplitude  of the stochastic
perturbation and  $\xi_j = \dfrac {dw_j}{dt}$ is a ``white noise" term that is the derivative of a Wiener
process. In matrix notation we can rewrite Eq. \ref{stochode} as
    \be
dx(t)=F(x(t))dt + \Sigma(x(t)) \, dw(t), \label{stochodem}
    \ee
where $\Sigma(x)=[\sigma_{ij}(x)]_{i,j=1,\ldots, d}$. Here it is always assumed that the It\^o, rather that the
Stratonovich, calculus, is used.  For a discussion of the differences see \cite{horsthemke84}, \cite{almcmbk94}
and \cite{risken84}. In particular, if the $\sigma_{ij}$ are independent of $x$ then the It\^o and  the
Stratonovich approaches yield identical results.

To fully interpret (\ref{stochodem})  we briefly review the properties of the $\xi$
when they are derived from a Wiener process.  We say that a continuous process $\{w(t)\}_{t>0}$ is a one
dimensional {\it Wiener process} if:
    \ben
    \item $w(0)=0$; and
    \item For all values of $s$ and $t$, $0 \leq s \leq t$ the random variable $w(t)-w(s)$ has the {\it Gaussian density}
    \be
    g(t-s,x) = \dfrac {1}{\sqrt{2\pi(t-s)}} \exp \left [-\dfrac {x^2}{2(t-s)}\right].
    \label {gaussian} 
    \ee
    \een
In a completely natural manner this  definition can be extended to say that the $d$ dimensional vector
    $$
    w(t) = \{w_1(t), \cdots, w_d(t)\}_{t>0}
    $$
is a {\it d-dimensional  Wiener process} if its components are one dimensional Wiener processes. Because of the
independence of the increments, it is elementary that the joint density of $w(t)$ is
    \be
    g(t,x_1,\cdots,x_d)=g(t,x_1)\cdots g(t,x_d)
    \label{g}
    \ee
and thus
    \be
    \int_{R^d} g(t,x)\,dx = 1,
    \label{normal}
    \ee
that
    \be
    \int_{R^d} x_ig(t,x)\,dx = 0, \qquad i=1,\cdots,d
    \label{aver}
    \ee
and
    \be
    \int_{R^d} x_ix_jg(t,x)\,dx = \delta_{ij}t \qquad i,j=1 ,\cdots,d.
    \label{var}
    \ee
In (\ref{var}),
    \be
    \delta_{ij}= \left\{
    \begin{array}{ll}
    1 & i=j \\
    0 & i \neq j
    \end{array}
    \right.
    \label{delta}
    \ee
is the Kronecker delta. Therefore  the average of a Wiener variable is zero by Eq. \ref{aver}, while the variance
increases linearly with time according to (\ref{var}).

The {\it Fokker-Planck equation}  that governs the evolution of the density function $f(t,x)$ of the process
$x(t)$ generated by the solution to the stochastic differential equation (\ref{stochodem}) is given by
    \be
    \frac {\partial f}{\partial t} = -  \sum_{i=1}^d \frac{\partial
    [F_i(x)f]}{\partial x_i} +\frac 12 \sum_{i,j=1}^d \frac{\partial ^2
    [a_{ij}(x)f]}{\partial x_i \partial x_j}
    \label{fpeqn}
    \ee
where
$$
a_{ij}(x)=\sum_{k=1}^{d}\sigma_{ik}(x)\sigma_{jk}(x).
$$
If $k(t,x,x_0)$ is the fundamental solution of the Fokker-Planck equation, i.e.
for every $x_0$ the function $(t,x)\mapsto k(t,x,x_0)$ is a solution of the
Fokker-Planck equation with the initial condition $\delta(x-x_0)$, then the
general solution $f(t,x)$ of the Fokker-Planck equation (\ref{fpeqn}) with the
initial condition
$$
f(x,0)=f_0(x)
$$
is given by \be f(t,x)=\int k(t,x,x_0)f_0(x_0)\, dx_0.
\label{gensoln}
\ee 
From a  probabilistic point of view $k(t,x,x_0)$ is a
stochastic kernel (transition density) and describes the
probability of passing from the state $x_0$  at time $t=0$
to the state $x$ at a time $t$.
Define the Markov operators $P^t$ by 
    \be
    P^tf_0(x)=\int k(t,x,x_0)f_0(x_0)\, dx_0, \quad f_0\in L^1.
    \label{mo}
    \ee
Then $P^tf_0$ is the density of the solution $x(t)$ of Eq. \ref{stochodem} provided that $f_0$ is the density of
$x(0)$.


The steady state density  $f_*(x)$ is the stationary solution of the Fokker Planck Eq. (\ref{fpeqn}):
    \be
    -  \sum_{i=1}^d \frac{\partial
    [F_i(x)f]}{\partial x_i} +\frac 12 \sum_{i,j=1}^d \frac{\partial ^2
    [a_{ij}(x)f]}{\partial x_i \partial x_j} = 0.
    \label{ssfpeqn}
        \ee
In the specific case of $X=\realnos^d$ and $\mu$ equal to the Lebesque measure, we recover from Eq. \ref{d:conent}
the conditional entropy
    \be
    H_c(f(t)|f_*) = -\int_{\realnos^d} f(t,x) \ln \left [ \dfrac {f(t,x)}{f_*(x)} \right ]\,dx
    \label{condent}
    \ee

If the coefficients $a_{ij}$ and $F_i$ are sufficiently regular so that a
fundamental solution $k$ exists, and $\int_X k(t,x,y)\, dx =1$, then the unique
generalized solution (\ref{gensoln}) to the Fokker-Planck equation
(\ref{fpeqn}) is given by Eq. \ref{mo}. One such set of  conditions is the
following: (1) the $F_i$ are of class $C^2$ with bounded derivatives; (2) the
$a_{ij}$ are of class $C^3$ and bounded with all derivatives bounded; and (3)
the uniform parabolicity condition holds, {\it i.e.} there exists a strictly
positive constant $\rho>0$ such that
$$
\sum_{i,j=1}^da_{ij}(x)\lambda_i\lambda_j\ge \rho \sum_{i=1}^d \lambda_i^2,\qquad \lambda_i,\lambda_j\in \realnos,
x\in \realnos^d.
$$
The uniform parabolicity condition implies that $k(t,x,y)>0$, and thus $P^tf(x)>0$ for every density, which
implies that there can be at most one stationary density, and that  if it exists then $f_*>0$. In this setting,
the corresponding conditional entropy $H_c(P^tf_0|f_*)$ approaches its maximal value of zero, as $t \rightarrow
\infty$, if and only if there is a stationary density $f_*$  that satisfies (\ref{ssfpeqn}).

\subsection{The one dimensional case}\label{ss:1d}

If we are dealing with a one  dimensional system, $d=1$, then the stochastic differential Eq. \ref{stochode}
 simply becomes
    \be
    \dfrac{dx}{dt} = F(x) +\sigma (x) \xi,
    \label{1dstochode}
    \ee
where again $\xi$ is a (Gaussian  distributed) perturbation with zero mean and
unit variance, and $\sigma (x)$ is the amplitude of the perturbation. The
corresponding Fokker-Planck equation (\ref{fpeqn}) becomes
    \be
    \dfrac{\partial f}{\partial t}
    = - \dfrac {\partial [F(x)f]}{\partial x}
    + \dfrac 12 \dfrac {\partial^2 [\sigma^2(x)f]}{\partial x^2}.
    \label{1dfpeqn}
    \ee
The Fokker-Planck equation  can also be written in the equivalent
form
    \be
    \dfrac{\partial f}{\partial t} = - \dfrac {\partial S}{\partial x}
    \label{fpcurrent}
    \ee
where
    \be
    S = -\dfrac12 \dfrac {\partial [\sigma^2(x)f]}{\partial x} + F(x)f
    \label{current}
    \ee
is called the {\it probability current.}

When stationary solutions of  (\ref{1dfpeqn}), denoted by $f_*(x)$ and defined
by $P_tf_*=f_*$ for all $t$, exist they are given as the generally unique (up
to a multiplicative constant) solution of (\ref{ssfpeqn}) in the case $d=1$:
    \be
    - \dfrac {\partial [F(x)f_*]}{\partial x} +
    \dfrac 12 \dfrac {\partial^2 [\sigma^2(x)f_*]}{\partial x^2} = 0.
    \label{fpstatden}
    \ee
Integration of Eq. \ref{fpstatden} by parts with the assumption that the probability current $S$ vanishes at the
integration limits, followed by a second integration, yields the solution
    \be
    f_*(x) = \dfrac {K}{\sigma^2(x)} \exp \left[ \int ^x \dfrac {2F(z)}{\sigma ^2(z)} \,dz \right].
    \label{statden}
    \ee
This stationary solution  $f_*$ will be a density if and only if
there exists a positive constant $K>0$ such that $f_*$ can be
normalized.

We now discuss rigorous results concerning the one dimensional case. Let $a(x)=\sigma^2(x)$. Assume that $a, a',$
and $F$ are continuous on $X=(\alpha,\beta)$, where $\infty\le \alpha<\beta\le\infty$. Let $a(x)>0$ for all $x\in
(\alpha,\beta)$ and $x_0$ be any point in $(\alpha,\beta)$. If we want to study the long term behaviour, as
$t\to\infty$, of the process $x(t)$ given by Eq. \ref{1dstochode}, we need to know that the process exists for all
$t>0$; in other words that there is no explosion in finite time, and that it lives in the interval
$(\alpha,\beta)$. If, for example, $\alpha$ were absorbing so that with positive probability we could reach
$\alpha$ in finite time, then a solution $f$ of Eq. \ref{1dfpeqn} would show a decrease in norm in
$L^1(\alpha,\beta)$.

There is a relation between the behaviour of $x(t)$ at the
boundary points $\alpha$ and $\beta$, and the existence and
uniqueness of solutions of the corresponding Fokker-Planck
equation as described by \cite{feller52, feller54}. In particular,
if the condition
    \begin{widetext}
    \be
    \int_\alpha^{x_0} \exp \left[- \int_{x_0} ^x \dfrac {2F(z)}{\sigma^2(z)}
    \,dz\right]
    \,dx=\int_{x_0}^\beta\exp \left[- \int_{x_0} ^x \dfrac {2F(z)}{\sigma^2(z)} \,dz
    \right]\,dx = \infty\label{e:reccurence}
    \ee
    \end{widetext}
holds, then the generalized solutions $P^tf$ of the Fokker-Planck equation exist and constitute a semigroup of
Markov operators on $L^1(\alpha,\beta)$. If only one of the integrals is finite then this conclusion holds as
well, but we restrict ourselves to Condition \ref{e:reccurence} because it is also necessary for the existence
of a stationary solution of Eq. \ref{1dstochode}, c.f. \citet{pinsky95}.  Thus we conclude that there is a
stationary density $f_*$ if and only if Condition \ref{e:reccurence} holds and
    \be
    \int_\alpha^{\beta} \frac{1}{\sigma^2(x)}\exp \left[ \int_{x_0} ^x \dfrac {2F(z)}{\sigma^2(z)}
    \,dz\right]\, dx<\infty.
    \label{1dsd}
    \ee

%
%

\cite{arnold01} provide a general method of establishing exponential convergence to zero of the relative
entropy for the Fokker-Planck equation in a divergence form. Here, we adapt the arguments of \cite{arnold01}
to the one-dimensional case with $X=(\alpha,\beta)$. Define
$$
D(x)=\dfrac{\sigma^2(x)}{2} \quad\mbox{and}\quad B(x)=-\ln \dfrac{K}{\sigma^2(x)}-\int_{0}^x
\dfrac{2F(z)}{\sigma^2(z)}\, dz,
$$
where $K$ is such that $f_*(x)=e^{-B(x)}$ is normalized. Then the Fokker-Planck equation \ref{1dfpeqn} can be
rewritten as
    \be
\dfrac{\partial f}{\partial t}
    =  \dfrac {\partial }{\partial
    x}\left(D(x)\left(\dfrac {\partial f}{\partial
    x}+ B'(x)f\right)\right)
    \label{1dfpeqdiv}
    \ee
and
$$
f_*(x)=e^{-B(x)}.
$$
We are going to show that if there exists a constant $\lambda>0$
such that for every $x\in X$ the following inequality holds
    \be
    \dfrac 12 D(x)B''(x)+D'(x)B'(x)+\dfrac{(D'(x))^2}{4 D(x)}-\dfrac 12 D''(x)
    \ge \lambda
    \label{e:1dexpbnd}
    \ee
then
    \be
    H_c(P^tf_0|f_*)\ge e^{-2\lambda t}H_c(f_0|f_*) \label{e:1dexpcon}
    \ee
for all initial densities $f_0$ with $H_c(f_0|f_*)>-\infty$.

The method of proof of Inequality \ref{e:1dexpcon} exploits Eq. \ref{e:1dexpbnd}. From Eq. \ref{1dfpeqdiv} it
follows that
    \be
\dfrac{\partial f}{\partial t}=\dfrac {\partial }{\partial
    x}\left(De^{-B} g\right)
    \label{1dfpeqsimp}
    \ee
where
$$
g=\dfrac{\partial }{\partial x}\left(e^B f\right).
$$
Let $\psi(z)=-z\log z$ and $H_c(t)=H_c(f|f_*)$, where
$f(t,x)=P^tf_0(x)$. Then we have
$$
H_c(t)=\int_X \psi(e^B f)e^{-B}dx.
$$
Differentiate $H_c$ with respect to $t$
$$
\dfrac{dH_c(t)}{dt}=\int_X \psi'(e^B f)\dfrac{\partial f}{\partial
t}\,dx,
$$
use Eq. \ref{1dfpeqsimp}, and integrate by parts  to obtain
$$
\dfrac{dH_c(t)}{dt}=-\int_X \psi''(e^B f)De^{-B} g^2 \, dx.
$$
Now write $I(t)=\dfrac{dH_c(t)}{dt}$ and differentiate $I(t)$
\begin{widetext}
    $$
\dfrac{dI(t)}{dt}=-\int_X \psi'''(e^B f)D g^2 \dfrac{\partial
f}{\partial t}\, dx -2\int_X \psi''(e^B f)D e^{-B} g \dfrac{\partial
g}{\partial t}\, dx
    $$
    \end{widetext}
Then use Eq. \ref{1dfpeqsimp}, rearrange terms and integrate by parts to obtain \citep{arnold01}
\begin{widetext}
$$
\dfrac{dI(t)}{dt}=-2\int_X \psi''(e^B f)D e^{-B}g^2\left(\dfrac 12 D''- D'B'-DB''-\dfrac{(D')^2}{4 D}\right)\,
dx+\int_X \Tr(Y\cdot Z)e^{-B}dx
$$
\end{widetext} where
$$
Y=\left(%
\begin{array}{cc}
  2\psi''(e^B f) & 2\psi'''(e^B f) \\
  2\psi'''(e^B f) & \psi'^{v}(e^B f) \\
\end{array}%
\right)
$$
 and
 $$
 Z=\left(%
\begin{array}{cc}
  \left(D\dfrac{\partial g}{\partial x}+\dfrac 12 D'g\right)^2 & D^2\dfrac{\partial g}{\partial x}g^2 +\dfrac 12 DD'g^3 \\
  D^2\dfrac{\partial g}{\partial x}g^2+ \dfrac 12 DD'g^3 & g^4 \\
\end{array}%
\right).
$$
Since $-\psi''(e^B f)D e^{-B}g^2$ is nonnegative, from Eq. \ref{e:1dexpbnd} it follows that the first integral
is less then $2\lambda \int_X \psi''(e^B f)D e^{-B}g^2 \,dx=-2\lambda I(t)$. Since $Y$ is a negative definite
matrix and $Z$ is a nonnegative definite matrix, we obtain $\Tr(Y\cdot Z)\le 0$
which completes the proof of the inequality
%
$$
\dfrac{dI(t)}{dt}\le -2\lambda I(t)
$$
Integrating from $t=s$ to $t=\infty$ and noting that $I(t)$ and $H_c(P^tf_0|f_*)$ go to zero as $t\to\infty$, we
arrive at
$$
- I(s)\le 2\lambda H_c(s),
$$
which leads to
$$
\dfrac{dH_c(s)}{ds}\ge -2\lambda H_c(s),
$$
and finally to
$$
H_c(t)\ge e^{-2\lambda t}H_c(0).
$$

\subsubsection{Example of an Ornstein-Uhlenbeck process}\label{sss:1dou}

Trying to find specific  examples of $\sigma(x)$ and $F(x)$ for
which one can determine the {\it time dependent}
 solution $f(t,x)$ of Eq.
\ref{1dfpeqn} is not easy.   
One solution that is known  is the one for  an Ornstein-Uhlenbeck process.  Since it is an Ornstein-Uhlenbeck
process, which was historically developed in thinking about perturbations to the velocity of a Brownian
particle, we denote the dependent variable by $v$ so we have $\sigma(v) \equiv \sigma$ a constant, and
$F(v) = -\gamma v$ with $\gamma \geq 0$. In this case, Eq. \ref{1dstochode} becomes
    \be
    \dfrac{dv}{dt} = -\gamma v  +\sigma \xi
    \label{o-ueqn}
    \ee
with the corresponding Fokker-Planck equation \be
    \dfrac{\partial f}{\partial t}
    =  \dfrac {\partial [\gamma v f]}{\partial v}
    + \dfrac {\sigma^2}2 \dfrac {\partial^2 f}{\partial v^2}.
    \label{o-ufpeqn}
    \ee
The unique stationary solution is
    \be
    f_*(v) =  \dfrac {e^{-\gamma v^2/\sigma^2}}{\int_{-\infty}^{+\infty}
    e^{-\gamma v^2/\sigma^2}dv}=  \sqrt{\dfrac {\gamma}{\pi \sigma^2}}e^{-\gamma v^2/\sigma^2}.
    \label{o-ustat}
    \ee
Further, from  \citet[Eq. 5.28]{risken84} the fundamental solution $k(t,v,v_0)$ is given by
    \be
    k(t,v,v_0)=\frac{1}{\sqrt{2 \pi b(t)}}\exp \left \{-\frac{(v-\exp(-\gamma
    t)v_0)^2}{2b(t)}\right \},
    \label{o-utdep}
    \ee
where
    \be
    b(t)= \frac{\sigma^2}{2\gamma}\left [1- e^{-2\gamma t} \right ].
    \label{b}
    \ee
Note that for a given $y$ and $t$  the function $k(t,\cdot,y)$ is a Gaussian
density with mean $v_0\exp(-\gamma t)$ and variance $b(t)$. Since
$k(t,v,v_0)>0$ for all $v,v_0\in\realnos$, the semigroup of Markov operators
$$
P^tf_0(v)=\int_{\realnos}k(t,v,v_0)f_0(v_0)\, dv_0
$$
satisfies all of the conditions of Theorem \ref{t:unique}. Thus we can assert
from Theorem \ref{t:unique} that $\lim_{t \to \infty}H_c(P^tf_0|f_*) = 0$ for
all $f_0$ with $H_c(f_0|f_*)>-\infty$.

In the case of the Ornstein-Uhlenbeck process, the sufficient condition (\ref{e:1dexpbnd}) for the exponential
lower bound on the conditional entropy reduces to
$$
D B''(v)\ge \lambda
$$
where
$$
D=\dfrac{\sigma^2}{2}\quad\mbox{and}\quad B''(v)=\dfrac{2\gamma}{\sigma^2}.
$$
Thus $\lambda=\gamma$ and Eq. \ref{e:1dexpcon} becomes
$$
H_c(P^tf_0|f_*)\ge e^{-2\gamma t}H_c(f_0|f_*)
$$
for all initial densities $f_0$ with $H_c(f_0|f_*)>-\infty$.

We will now show that this lower bound is optimal. Let us first calculate the conditional entropy of two Gaussian
densities. Let $q_1,q_2>0$ and let $z_1,z_2\in \realnos$. Consider densities of the form
$$
g_i(x,z_i)=\sqrt{\dfrac{q_i}{\pi}}\exp\{- q_i(x-z_i)^2\},\qquad x\in \realnos, \quad i=1,2.
$$
Then
$$
\log \dfrac{g_1(x,z_1)}{g_2(x,z_2)}=\log \sqrt{\dfrac{q_1}{q_2}}-q_1(x-z_1)^2+q_2(x-z_2)^2.
$$
Since
$$
\int_{\realnos} g_1(x,z_1)x^2\, dx=\dfrac{1}{2q_1}+z_1^2\quad\mbox{and}\quad \int_{\realnos}g_1(x,z_1)x\,
dx=z_1,
$$
we arrive at
    \be
H_c(g_1(\cdot,z_1)|g_2(\cdot,z_2))=\dfrac 12 \log
\dfrac{q_2}{q_1}+\dfrac{1}{2}\left(1-\dfrac{q_2}{q_1}\right)-q_2(z_1-z_2)^2. \label{e:conent1dg}
    \ee

Now let $f_0$ be a Gaussian density of the form
$$
f_0(v)=\sqrt{\dfrac{c_1}{\pi}}\exp\{-c_1 (v-c_2)^2\},
$$
where $c_1>0$ and $c_2\in\realnos$. Since
$$
P^tf_0(v)=\int_{\realnos } k(t,v,v_0)f_0(v_0)\, dv_0,
$$
we obtain by direct calculation using Eq. \ref{o-utdep}
\begin{widetext}
$$
P^tf_0(v)=\sqrt{\dfrac{c_1}{\pi(e^{-2\gamma t}+2c_1 b(t))} }\exp\left\{-\dfrac{c_1}{e^{-2\gamma t}+2c_1
b(t)}\left(v-c_2 e^{-\gamma t}\right)^2\right\}.
$$
\end{widetext}
Consequently, from Eq. \ref{e:conent1dg}, with $q_1=\dfrac{c_1}{e^{-2\gamma t}+2c_1 b(t)}$,
$q_2=\dfrac{\gamma}{\sigma^2}$, $z_1=c_2 e^{-\gamma t}$, and $z_2=0$, it follows that
\begin{widetext}
    \be
    H_c(P^tf_0|f_*)=\dfrac 12 \log\left [1-\left(1-\dfrac{\gamma}{\sigma^2c_1}\right)e^{-2\gamma
    t}\right ] +\dfrac{1}{2}\left(1-\dfrac{\gamma}{\sigma^2c_1}-\dfrac{2\gamma c_2^2}{\sigma^2}\right)e^{-2\gamma
    t}.
    \ee
\end{widetext}
In particular, if $c_1=\dfrac{\gamma}{\sigma^2}$ then
$
H_c(f_0|f_*)=-\dfrac{\gamma c_2^2}{\sigma^2}
$
and
$$
H_c(P^tf_0|f_*)=e^{-2\gamma t} H_c(f_0|f_*),\qquad t\ge 0.
$$

\subsubsection{A Rayleigh process}\label{sss:rp}

Another  example for which we have an analytic form for the density
$f(t,x)$ is that of a {\it Rayleigh process} \citep[page
135-136]{gardinerhandbook}.  In this case we have
    \be
    \dfrac{dv}{dt} = -\gamma v + \dfrac{\sigma ^2}{2v} + \sigma \xi \qquad v \in [0, \infty)
    \label{rayleigh}
    \ee
and the associated Fokker-Planck equation is
    \be
    \dfrac{\partial f}{\partial t}
    =  \dfrac {\partial }{\partial v} \left [ \left ( \gamma v - \dfrac{\sigma ^2}{2v} \right ) f\right ]
    + \dfrac {\sigma^2}2 \dfrac {\partial^2 f}{\partial v^2}.
    \label{ray-fpeqn}
    \ee
The unique equilibrium  solution, with the proper normalization, is given by
    \be
    f_*(v) = \dfrac {2 \gamma v }{\sigma ^2} e^{-\gamma v^2 /\sigma^2}
    \label{ray-equil}
    \ee
and the fundamental solution is
    \be
    k(t,v,v_0) =\dfrac {2 \gamma v }{\sigma ^2} e^{-\gamma v^2 /\sigma^2}
    \sum_{n=0}^\infty L_n \left ( \dfrac {\gamma v_0^2}{\sigma^2} \right)
    L_n \left ( \dfrac {\gamma v^2}{\sigma^2} \right)
    e^{-2n\gamma t},
    \label{ray-f}
    \ee
where the $L_n$ are the Laguerre polynomials of order $n$. This form of $k$ is not particularly useful in
calculations. The Rayleigh process is also known as a  two-dimensional radial Ornstein-Uhlenbeck
process\footnote{Given two independent Ornstein-Uhlenbeck processes $v_1(t)$ and $v_2(t)$ with parameters
$\gamma$ and $\sigma$ as defined in Eq. \ref{o-ueqn}, the two-dimensional radial Ornstein-Uhlenbeck process is
then defined as $\sqrt{v_1(t)^2+v_2(t)^2}.$} with parameters $\gamma$ and $\sigma$, and as such its transition
probability  is given by \citep{borodin}
    \be
    k(t,v,v_0)=\dfrac{v}{b(t)}e^{ -(v^2+e^{-2\gamma
    t}v_0^2)/(2b(t))} I_0\left(\dfrac{v v_0}{e^{\gamma
    t}b(t)}\right)
    \ee
where $b(t)=\dfrac{\sigma^2}{2\gamma}(1-e^{-2\gamma t})$ and $I_0(z)=\sum_{k=0}^\infty\dfrac{z^{2k}}{(2^kk!)^2}$
is the modified Bessel function of the first kind. Since $k(t,v,v_0)>0$ for all $t,v,v_0>0$, again from Theorem
\ref{t:unique} it follows that $\lim_{t \to \infty}H_c(P^tf_0|f_*) = 0$ for all $f_0$ with
$H_c(f_0|f_*)>-\infty$. 

In the case of the Rayleigh process, the sufficient Condition \ref{e:1dexpbnd} for the exponential lower bound
on the conditional entropy reduces to
$$
D B''(v)\ge \lambda
$$
where
$$
D=\dfrac{\sigma^2}{2}\quad\mbox{and}\quad B''(v)=\dfrac{2\gamma}{\sigma^2}+\dfrac{1}{v^2}.
$$
Thus $\lambda=\gamma$ and Eq. \ref{e:1dexpcon} becomes
$$
H_c(P^tf_0|f_*)\ge e^{-2\gamma t}H_c(f_0|f_*)
$$
for all initial densities $f_0$ with $H_c(f_0|f_*)>-\infty$.

\subsection{The multidimensional case }
\label{ss:md}

In this section first we turn our attention to the existence of a stationary density in the case of
multidimensional diffusion when the matrix $\Sigma(x)$ is nonsingular at every point $x$. This case is much more
involved and does not yield simple necessary and sufficient conditions for the existence of a stationary
density.  However, when we consider two specific examples much more can be said.

Let us assume that $F$ and $a$ are of class $C^2$ and $C^3$ respectively and that
$$
\sum_{i,j=1}^d a_{ij}(x)\lambda_i\lambda_j>0,\quad \mbox{for}\quad x\in\realnos^d\quad \mbox{and}\quad
\lambda\in \realnos^d-\{0\}.
$$
Under these assumptions, the so-called Liapunov method \citep{pinsky95} implies that if there exists a
nonnegative function $V$ of class $C^2$ on $\realnos^d$ and a positive constant $r>0$ such that
    \be
    \sup_{|x|\ge r}L^- V(x)<0, \label{e:liapunov}
    \ee
where the operator
    \be
    L^-=  \sum_{i=1}^d F_i(x) \frac{\partial }{\partial x_i} +\frac
    {1}{2} \sum_{i,j=1}^d a_{ij}(x) \frac{\partial ^2 }{\partial x_i \partial
    x_j},
    \label{l-}
    \ee
then there is a unique stationary density $f_*$ and $f_*(x)>0$ for all $x$.
Again, from Theorem \ref{t:unique} $\lim_{t \to \infty}H_c(P^tf_0|f_*) = 0$ for
all $f_0$ with $H_c(f_0|f_*)>-\infty$. If we simply take $V(x)=|x|^2$, then
Condition \ref{e:liapunov} becomes
$$
\sup_{|x|\ge r}\left (\sum_{i,j=1}^d a_{ij}(x)+ 2\sum_{i=1}^dx_iF_i(x)
\right)<0.
$$

One can also write an integral test \citep{bhattacharya78} which reduces to Condition \ref{1dsd} in the
one-dimensional case. Let
$$
A_1(x)=\frac{1}{|x|^2} x^T a(x) x=\frac{1}{|x|^2}\sum_{i,j=1}^d a_{ij}(x)x_ix_j,
$$
and
$$
A_2(x)=\Tr(a(x)) +2\langle x, F(x)\rangle=\sum_{i=1}^d a_{ii}(x) + 2\sum_{i=1}^d x_i F_i(x).
$$
Now define
$$
{\beta}(r)=\sup_{|x|=r}\frac{A_2(x)-A_1(x)}{A_1(x)}\quad\mbox{and} \quad \alpha(r)=\inf_{|x|=r}A_1(x).
$$
Finally, for $r\ge 1$, let
    $$
    {J}(r)=\int_{1}^r\frac{{\beta}(s)}{s}\, ds
    $$
and assume that there exists $r_0>0$ such that the following two conditions hold:
    \be
    \int_{r_0}^\infty
    e^{-J(r)}\, dr =\infty\quad\mbox{and}\quad
    \int_{r_0}^\infty\frac{1}{\alpha(r)} e^{J(r)}\, dr
    <\infty.\label{e:idmulti}
    \ee
We claim that Condition \ref{e:idmulti} implies Condition \ref{e:liapunov}. To see this, for $r\ge r_0$ define
$$
\tilde{J}(r)=\int_1^re^{-J(s)}\int_s^\infty \frac{1}{\alpha(q)} e^{J(q)}\, dq\,ds,
$$
and take $V$ to be of class $C^2$ on $\realnos^d$ such that $V(x)=\tilde{J}(|x|)$ for $|x|\ge r_0$. It is easy
to check that
$$
2L^-V(x)= A_1(x) \tilde{J}''(|x|)+\frac{\tilde{J}'(|x|)}{|x|}(A_2(x)-A_1(x))
$$
and that $2L^-V(x)\le -1$ for $|x|\ge r_0$.

We now discuss the speed of convergence of the conditional entropy. \cite{arnold01} provide a general sufficient
condition for such convergence in the case of multidimensional diffusions 
that the Fokker-Planck equation can be written in a divergence form:
\be
    \frac {\partial f}{\partial t} = \mathrm{div}\left(D(x)(\nabla f+ f\nabla
    B(x))\right),
    \label{fpeqndiv0}
    \ee
where $D(x)$ is locally uniformly positive and $B(x)$ is a real-valued function such that the stationary
solution
$$
f_*(x)=e^{-B(x)}
$$
is a density. For simplicity assume that $D(x)=D$ is a constant. Then the Fokker-Planck equation \ref{fpeqndiv0}
describes the evolution of densities for the system
    \be
    dx(t)=-D\nabla B(x(t))dt+\sqrt{2 D}I dw(t),
    \ee
where $I$ is the identity matrix.
If there is $\lambda>0$ such that
    \be
    D\left(\frac{\partial ^2
    B(x)}{\partial x_i \partial x_j}\right)_{i,j=1,\ldots d}\ge
    \lambda \,\mathrm{I},\label{e:con}
    \ee
then we have
    \be
H_c(P^tf_0|f_*)\ge e^{-2\lambda t}H_c(f_0|f_*). \label{e:expcon}
    \ee
for all initial densities $f_0$ with $H_c(f_0|f_*)>-\infty$.

\subsection{Multidimensional Ornstein-Uhlenbeck process}\label{ss:mo-up}

Consider the multidimensional Ornstein-Uhlenbeck process
    \be
    \dfrac{dx}{dt} = Fx  +\Sigma \xi
    \label{linear}
    \ee
where $F$ is a $d\times d$ matrix, $\Sigma$ is a $d\times d$ matrix an $\xi$ is $d$ dimensional vector. The formal
solution to Eq. \ref{linear} is given by
    \be
    x(t)=e^{tF}x(0)+\int_0^t e^{(t-s)F}\Sigma \, dw(t)
    \label{slinear}
    \ee
where $e^{tF}=\sum_{n=0}^\infty \frac{t^n}{n!}F^n$ is the
fundamental solution to $\dot{X}(t)=FX(t)$ with $X(0)=I$, and $w(t)$
is the standard $d$-dimensional Wiener process. From the properties
of stochastic integrals it follows that
$$
\eta(t)=\int_0^t e^{(t-s)F}\Sigma\, dw(t)
$$
has mean $0$ and covariance
$$
R(t)=E\eta(t)\eta(t)^T=\int_{0}^t e^{sF}\Sigma\Sigma^T e^{sF^T} ds,
$$
where $B^T$ is the transpose of the matrix $B$. The matrix $R(t)$ is nonnegative definite but not necessarily
positive definite. We follow the presentation of \cite{zakaisnyders70} and \cite{erickson71}. For each $t>0$ the
matrix $R(t)$ has constant rank equal to the dimension of the space
$$
[F,\Sigma]:=\{F^{l-1}\Sigma \epsilon_j: l,j=1,\ldots,d,
\epsilon_j=(\delta_{j1},\ldots,\delta_{jp})^T\}.
$$
If $m=\rank R(t)$ then $d-m$ coordinates of the process $\eta(t)$ are equal to $0$ and the remaining $m$
coordinates constitute an $m$-dimensional Gaussian process. Thus if $m<d$ there is no stationary density. If
$\rank R(t)=d$ then the transition probability function of $x(t)$ is given by the Gaussian density
\begin{widetext}
    \be
    k(t,x,x_0)=\frac{1}{(2\pi)^{d/2}(\det
    R(t))^{1/2}}\exp\{-\frac{1}{
    2}(x-e^{tF}x_0)^TR(t)^{-1}(x-e^{tF}x_0)\},
    \label{e:tdmo-up}
    \ee
\end{widetext}
where $R(t)^{-1}$ is the inverse matrix of $R(t)$. An invariant density $f_*$ exists if and only if all
eigenvalues of $F$ have negative real parts, and in this case the unique stationary density $f_*$ has the form
    \be
    f_*(x)=\frac{1}{(2\pi)^{d/2}(\det
    R_{*})^{1/2}}\exp\left \{-\frac{1}{2}x^T R_{*}^{-1}x\right \},
    \label{e:sdmo-up}
    \ee
where $R_{*}$ is a positive definite matrix given by
    $$
    R_{*}=\int_0^\infty e^{sF}\Sigma\Sigma^T e^{sF^T} ds,
    $$
    and is a unique symmetric matrix satisfying
    \be
    F R_*+R_*F^T=-\Sigma\Sigma^T.\label{e:Leq}
    \ee
We conclude that if $[F,\Sigma]$ contains $d$ linearly independent vectors and all eigenvalues of $F$ have
negative real parts, then from Theorem \ref{t:unique} it follows that $\lim_{t \to \infty}H_c(P^tf_0|f_*) = 0$
for all $f_0$ with $H_c(f_0|f_*)>-\infty$. 


To simplify the following we first recall some properties of multivariate Gaussian distributions.
Let $Q_1$, $Q_2$ be positive definite matrices and let
$$
g_i(x,z)=\frac{1}{(2\pi)^{d/2}(\det
    Q_i)^{1/2}}\exp\left \{-\frac{1}{2}(x-z)^T Q_i^{-1}(x-z)\right \}.
$$
Then
\begin{widetext}
    \be
    \log \frac{g_1(x,z_1)}{g_2(x,z_2)}=\frac{1}{2}\log \frac{\det
    Q_2}{\det Q_1}-\frac{1}{2}(x-z_1)^T
    Q_1^{-1}(x-z_1)+\frac{1}{2}(x-z_2)^T Q_2^{-1}(x-z_2).
    \label{e:logmulti}
    \ee
\end{widetext}
Since $\int g_1(x,0)x x^T\, dx=Q_1$, we have
$$
\int g_1(x,z_1)x x^T\, dx=Q_1+z_1 z_1^T\quad\mbox{and}\quad \int
g_1(x,z_1)x\,dx=z_1.
$$
Note also that $z^T Q z$ can be written with the help of the trace
of a matrix as $\Tr[Q z z^T]$ for any matrix $Q$ and any vector $z$.
Consequently,
\begin{widetext}
    \be
    H_c(g_1(\cdot, z_1)|g_2(\cdot, z_2))=\frac{1}{2}\log \frac{\det
    Q_1}{\det Q_2}+\frac{1}{2}\Tr [(Q_1^{-1}-
    Q_2^{-1})Q_1]-\frac{1}{2}\Tr[Q_2^{-1}(z_1-z_2)(z_1-z_2)^T].
    \label{e:entropyd}
    \ee
\end{widetext}

Now let $f_0$ be a Gaussian density of the form
\begin{widetext}
\be
f_0(x)=\frac{1}{(2\pi)^{d/2}(\det
    V(0))^{1/2}}\exp\left \{-\frac{1}{2}(x-m(0))^T V(0)^{-1}(x-m(0))\right \}\label{e:gaussini}
\ee
\end{widetext}
where $V(0)$ is a positive definite matrix and $m(0)\in\realnos^d$. From Eq. \ref{slinear} it follows that
$x(t)$ is Gaussian with the following mean vector $ m(t)=e^{tF}m(0)$ and  covariance matrix
   $V(t)=e^{tF}V(0) e^{tF^T}+ R(t)$.
Hence
\begin{widetext}
$$
P^tf_0(x)=\frac{1}{(2\pi)^{d/2}(\det
    V(t))^{1/2}}\exp\left \{-\frac{1}{2}(x-m(t))^T V(t)^{-1}(x-m(t))\right \}
$$
\end{widetext}
Consequently, from Eq. \ref{e:entropyd}, with $Q_1=V(t)$, $z_1=m(t)$, $Q_2=R_{*}$, and $z_2=0$, we
obtain
\begin{widetext}
    \be
H_c(P^tf_0|f_*)=\dfrac 12 \log\dfrac{\det V(t)}{\det R_{*}}+\dfrac 12 \Tr(\mathrm{I}-R_*^{-1}V(t))-\dfrac 12
\Tr(R_{*}^{-1}m(t)m(t)^T).\label{f:cemou1}
    \ee
\end{widetext} In particular, if $V(0)=R_*$, then $V(t)=R_*$ and
    \be
H_c(P^tf_0|f_*)=-\dfrac 12 \Tr(R_{*}^{-1}m(t)m(t)^T)\label{f:cemou2}
    \ee
for all $t\ge 0$ and every $f_0$ of the form given by Eq. \ref{e:gaussini}.
%

As a specific example of the multidimensional Ornstein-Uhlenbeck process, to which Equation \ref{e:expcon} and
\ref{f:cemou2} can be applied, consider the case when $\Sigma=\sigma\mathrm{I}$ and $F$ is a diagonal matrix,
$F=-\lambda \mathrm{I}$ with $\lambda>0$. Then $R_*^{-1}=\dfrac{2\lambda}{\sigma^2}\mathrm{I}$ and
$f_*(x)=e^{-B(x)}$, where
$$B(x)=\dfrac 12 \log ((2\pi)^{d}\det
    R_{*})+ \dfrac{\lambda}{\sigma^2} x^Tx.$$
Thus Condition \ref{e:con} becomes
$$
\dfrac {\sigma^2}{2}\left(\frac{\partial ^2
    B(x)}{\partial x_i \partial x_j}\right)_{i,j=1,\ldots d}=
    \lambda \,\mathrm{I}.
$$
Since $e^{tF}=e^{-\lambda t}\mathrm{I}$,  we conclude from Equation \ref{f:cemou2} that
$H_c(P^tf_0|f_*)=e^{-2\lambda t}H_c(f_0|f_*)$, so the estimate in Equation \ref{e:expcon} is optimal. We will
show in the next section that in the case of a non-invertible matrix $\Sigma$ a slower speed of convergence
might occur.

To obtain a lower bound on the conditional entropy for the case of a not necessarily invertible matrix $\Sigma$
and a general $f_0$, we make use of the following inequality, proved in the appendix,
\begin{widetext}
\be H_c(P^tf_0|f_*)\ge \int\int f_0(y_1)f_*(y_2)H_c(k(t,\cdot,y_1)|k(t,\cdot,y_2))dy_1dy_2.
\label{e:lowbound}\ee
\end{widetext}
From Eq. \ref{e:entropyd} with $Q_1=R(t)$, $Q_2=R(t)$,
$z_1=e^{tF}y_1$, and $z_2=e^{tF}y_1$ it follows that
\begin{widetext}
$$
H_c(k(t,\cdot,y_1)|k(t,\cdot,y_2))=-\dfrac{1}{2}\Tr[ R(t)^{-1}e^{tF}(y_1-y_2)(y_1-y_2)^Te^{tF^T}].
$$
\end{widetext}
Since $ \Tr[ R(t)^{-1}e^{tF}yy^Te^{tF^T}]\le ||y||^2||e^{tF^T}R(t)^{-1}e^{tF}||$ for any $y$ by the Schwartz
inequality, we obtain from Eq. \ref{e:lowbound}
\begin{widetext}
    \be
H_c(P^tf_1|f_*)\ge-\dfrac{1}{2} ||e^{tF}||^2||R(t)^{-1}||\int\int ||y_1-y_2||^2f_0(y_1)f_*(y_2)\,dy_1 dy_2.
\label{e:lowbnde}
    \ee
\end{widetext} Finally, observe that the norm of $R(t)^{-1}$ is bounded, because $R(t)^{-1}$ converges to
$R_*^{-1}$ as $t\to\infty$. Thus for sufficiently large $t$ we have
\begin{widetext}
    \be
H_c(P^tf_1|f_*)\ge-||e^{tF}||^2||R_*^{-1}||\int\int ||y_1-y_2||^2f_0(y_1)f_*(y_2)\,dy_1
dy_2.\label{e:lowbnde1}
    \ee
\end{widetext}

\subsubsection{Harmonic oscillator}\label{sss:ho}

Consider the second order system
    \be
    m\dfrac{d^2x}{dt^2}+ \gamma \dfrac{dx}{dt}+\omega^2 x=\sigma \xi
    \label{e:b-osc}
    \ee
with constant positive coefficients $m$, $\gamma$ and $\sigma$. Introduce the velocity $v=\dfrac{dx}{dt}$ as a new
variable. Then Eq. \ref{e:b-osc} is equivalent to the system
    \begin{eqnarray}
    \dfrac{dx}{dt} &=& v \\\label{e:b-osc1}
    m \dfrac{dv}{dt} &=& -\gamma v-\omega^2 x +\sigma \xi.,\label{e:b-osc2}
    \end{eqnarray}
and the corresponding Fokker-Planck equation is
    $$
    \dfrac{\partial f}{\partial t}
    = - \dfrac {\partial [v f]}{\partial x}+ \dfrac {1}{m} \dfrac {\partial [(\gamma v +\omega^2 x )f]}{\partial v}
    + \dfrac {\sigma^2}{2 m^2} \dfrac {\partial^2 f}{\partial v^2}.
       $$
%
%
We can use the results of Section \ref{ss:mo-up} in the two dimensional setting with
$$
F=\left(\begin{array}{cc} 0 & 1\\
-\dfrac{\omega^2}{m}& -\dfrac{\gamma}{m}
\end{array}\right)
\quad\mbox{and}\quad \Sigma=\left(\begin{array}{cc}0 & 0\\
0 & \dfrac{\sigma}{m}
\end{array}\right).
$$
Since
$$
[F,\Sigma]=\left\{ \left(
\begin{array}{c}
 0 \\
 0 \\
\end{array}
\right),
\left(
\begin{array}{c}
  0 \\
  \dfrac{\sigma}{m} \\
\end{array}%
\right),
\dfrac{\sigma}{m}\left(%
\begin{array}{c}
  1 \\
  -\dfrac{\gamma}{m} \\
\end{array}%
\right) \right\},
$$
the transition density function is given by Eq. \ref{e:tdmo-up}. The eigenvalues of $F$ are equal to
$$
\lambda_1=\frac{-\gamma+\sqrt{\gamma^2-4
m\omega^2}}{2m}\quad\mbox{and}\quad\lambda_2=\frac{-\gamma-\sqrt{\gamma^2-4
m\omega^2}}{2m},
$$
and are either negative real numbers when $\gamma^2\ge 4m\omega^2$ or complex numbers with negative real parts
when $\gamma^2< 4m \omega^2$.  Thus the stationary density is given by Eq. \ref{e:sdmo-up}. As is easily seen
$R_*$, being a solution to Eq. \ref{e:Leq}, is given by
$$
R_{*}=\left(\begin{array}{cc}
\dfrac{\sigma^2}{2\gamma\omega^2}& 0 \\
0 & \dfrac{\sigma^2}{2m\gamma}\end{array}\right).
$$
The inverse of the matrix $R_*$ is
$$
R_{*}^{-1}=\dfrac{2\gamma}{\sigma^2} \left(\begin{array}{cc}
\omega^2 & 0 \\
0 & m\end{array}\right)
$$
and the unique stationary density becomes
$$
f_*(x,v)=  \dfrac {\gamma \omega \sqrt m}{\pi \sigma ^2} e^{-\dfrac {\gamma }{\sigma ^2}[\omega ^2 x^2+m v^2 ]}.
$$
As in Section \ref{ss:mo-up} we conclude that  $\lim_{t \to \infty}H_c(P^tf_0|f_*) = 0$ for all $f_0$ with
$H_c(f_0|f_*)>-\infty$.

The bound on the temporal convergence  of $H_c(P^tf_0|f_*)$ to zero, as given by Equation \ref{e:lowbnde1},
is determined by
$||e^{tF}||^2$. Thus we are going to calculate $||e^{tF}||^2$ 
and to see the nature of the general formula (\ref{f:cemou2}) for the conditional entropy in the case of the
harmonic oscillator. 
First, consider  the case when $\lambda_1=\lambda_2$,  that is $\gamma^2=4m\omega^2$, and set
$$
\lambda=-\frac{\gamma}{2m}.
$$
Then we have
$$
F=\left(%
\begin{array}{cc}
  0 & 1 \\
  -\lambda^2 & 2\lambda \\
\end{array}%
\right) \quad\mbox{and}\quad
e^{tF}=e^{\lambda t}\left(%
\begin{array}{cc}
  1-\lambda t & t \\
  -\lambda^2 t & 1+\lambda t \\
\end{array}%
\right)
$$
and
$$
||e^{tF}||^2=e^{2\lambda t}(1+(3\lambda^2+1)t^2).
$$
If we take $m(0)=(1,\lambda)^T$
then $m(t)=e^{\lambda t}m(0)$ and Eq. \ref{f:cemou2} becomes
$$
H_c(P^tf_0|f_*)=e^{2\lambda t}H_c(f_0|f_*).
$$
But, if we take $m(0)=(1,0)^T$, then $m(t)=e^{\lambda t}(1-\lambda t,-\lambda^2 t)^T$ and
$$
H_c(P^tf_0|f_*)=e^{2\lambda t}\left((1-\lambda t)^2+\lambda^2 t^2\right)H_c(f_0|f_*).
$$

Now consider the case when $\lambda_1\neq \lambda_2$ are real and define, for $t\ge 0$,
    $$
    c_1(t)=\frac{\lambda_2 e^{\lambda_1 t}-\lambda_1 e^{\lambda_2
    t}}{\lambda_2-\lambda_1}
    \quad\mbox{and}\quad
    c_2(t)=\frac{e^{\lambda_2 t}-e^{\lambda_1
    t}}{\lambda_2-\lambda_1}.
    $$
Then
$$
e^{tF}=\left(\begin{array}{cc} c_1(t)& c_2(t)\\
c'_1(t) & c'_2(t)\end{array}\right)
$$
and
$$
||e^{tF}||^2=e^{2\lambda_1 t}+e^{2\lambda_2 t}+ (\lambda_1\lambda_2+1)^2c_2^2(t).
$$
If we take $m(0)=(1,\lambda_i)^T$, $i=1,2$ in Equation \ref{f:cemou2}, then $m(t)=e^{\lambda_i t}m(0)$ and
$$
H_c(P^tf_0|f_*)=e^{2\lambda_i t}H_c(f_0|f_*).
$$

Finally, consider the case of complex $\lambda_1,\lambda_2$ and let
$$
\alpha=-\dfrac{\gamma}{2m}\qquad \mbox{and}\qquad \beta=\dfrac{\sqrt{4 m\omega^2-\gamma^2}}{2m}.
$$
Then
$$
e^{tF}=\dfrac{e^{\alpha t}}{\beta}\left(%
\begin{array}{cc}
  \beta\cos(\beta t)-\alpha\sin(\beta t) & \sin(\beta t) \\
  -(\alpha^2+\beta^2)\sin(\beta t) & \beta\cos(\beta t)+\alpha\sin(\beta t), \\
\end{array}%
\right)
$$
and we have
\begin{widetext}
$$
||e^{tF}||^2=\dfrac{e^{2\alpha t}}{\beta^2}\left(2\beta^2\cos^2(\beta t)+(2\alpha^2
+(\alpha^2+\beta^2)^2+1)\sin^2(\beta t)\right).
$$
\end{widetext}

%
%

\subsubsection{Colored noise}{\label{sss:cn}

Consider the system
    \be
    \dfrac{dx}{dt}=-\alpha x + \eta
    \label{e:colnoise}
    \ee
where $\alpha>0$ and $\eta$ is the one dimensional
Ornstein-Uhlenbeck process with parameters $\gamma,\sigma>0$
$$
\dfrac{d\eta}{dt}=-\gamma \eta + \sigma \xi.
$$
Then Eq. \ref{e:colnoise} is equivalent to the system
    \begin{eqnarray}
    \dfrac{dx}{dt} &=& -\alpha x + v \\\label{e:colnoise1}
    \dfrac{dv}{dt} &=& -\gamma v +\sigma \xi,\label{e:colnoise2}
    \end{eqnarray}
and the corresponding  Fokker-Planck equation is
    $$
    \dfrac{\partial f}{\partial t}
    =  \dfrac {\partial [(\alpha x -v) f]}{\partial x}+ \dfrac {\partial [\gamma v f]}{\partial v}
    + \dfrac {\sigma^2}{2} \dfrac {\partial^2 f}{\partial v^2}.
       $$
We can use the results of Section \ref{ss:mo-up} in the two dimensional setting with
$$
F=\left(\begin{array}{cc} -\alpha & 1\\
0 & -\gamma
\end{array}\right)
\quad\mbox{and}\quad \Sigma=\left(\begin{array}{cc}0 & 0\\
0 & \sigma
\end{array}\right).
$$
Observe that
$$
[F,\Sigma]=\left\{
\left(%
\begin{array}{c}
  0 \\
  0 \\
\end{array}%
\right)
\left(%
\begin{array}{c}
  0 \\
  \sigma \\
\end{array}%
\right),
\left(%
\begin{array}{c}
  \sigma \\
  -\gamma\sigma \\
\end{array}%
\right) \right\}.
$$
The eigenvalues of $F$ are given by
$$
\lambda_1=-\alpha \quad\mbox{and}\quad \lambda_2=-\gamma,
$$
and are evidently negative. Finally,
$$
R_{*}=\dfrac{\sigma^2}{2 \gamma (\alpha+\gamma)}
\left(%
\begin{array}{cc}
    \dfrac{1}{\alpha}
   & 1  \\
   1
   & \alpha +\gamma \\
\end{array}%
\right) $$
and
$$
R_{*}^{-1}=\dfrac{2 \alpha (\alpha+\gamma)}{\sigma^2}
\left(%
\begin{array}{cc}
 \alpha +\gamma
   & -1  \\
   -1
   &\dfrac{1}{\alpha}\\
\end{array}%
\right).
$$
Thus the unique stationary density is equal to
\begin{widetext}
$$
f_*(x,v)=\dfrac{\sqrt{\alpha\gamma}
(\alpha+\gamma)}{\pi\sigma^2}\exp\left\{-\dfrac{\alpha+\gamma}{\sigma^2}\left( \alpha(\alpha+\beta)x^2-2\alpha x
v+v^2\right)\right\}.
$$
\end{widetext} As in Section \ref{ss:mo-up} we conclude that $\lim_{t \to \infty}H_c(P^tf_0|f_*) = 0$ for all
$f_0$ with $H_c(f_0|f_*)>-\infty$.

Finally, we are going to calculate $||e^{tF}||^2$ to see how the general formula \ref{f:cemou2} for the
conditional entropy bound looks in this special case. First consider the case when $\alpha\neq \gamma$. The
fundamental matrix is given by
$$
e^{tF}=\left(%
\begin{array}{cc}
  e^{-\alpha t} & \beta(t) \\
  0 & e^{-\gamma t} \\
\end{array}%
\right)\quad\mbox{with}\quad \beta(t)=\dfrac{1}{\alpha-\gamma} \left(e^{-\gamma t}-e^{-\alpha t}\right)
$$
and
$$
||e^{tF}||^2=e^{-2\gamma t}+e^{-2\alpha t}.
$$
If we take $m(0)=(1,0)^T$,  in Equation \ref{f:cemou2}, then $m(t)=e^{-\alpha t}m(0)$ and
$$
H_c(P^tf_0|f_*)=e^{-2\alpha t}H_c(f_0|f_*).
$$
Similarly, for  $m(0)=(1,\alpha-\gamma)^T$ we have $m(t)=e^{-\gamma t}m(0)$ and
$$
H_c(P^tf_0|f_*)=e^{-2\gamma t}H_c(f_0|f_*).
$$

Now consider the case when $\alpha=\gamma$. We have
$$
e^{tF}=\left(%
\begin{array}{cc}
  e^{-\gamma t} & te^{-\gamma t} \\
  0 & e^{-\gamma t} \\
\end{array}%
\right)\quad\mbox{and}\quad ||e^{tF}||^2=e^{-2\gamma t}+e^{-2\alpha t}.
$$
Equation \ref{f:cemou2} becomes for  $m(0)=(0,1)^T$ and $m(t)=e^{-\gamma t}(t,1)^T$
$$
H_c(P^tf_0|f_*)=e^{-2\gamma t}(2\gamma^2 t^2-2\gamma t+1)H_c(f_0|f_*).
$$

\setcounter{equation}{0}
\section{Markovian dichotomous noise}\label{s:mdn}

Another example where we can use our results is the case of dichotomous noise \citep[Section 8]{horsthemke84}.
The state space of the Markovian dichotomous noise $\xi(t)$ consists of two states $\{c_+,c_{-}\}$ and is
characterized by a transition probability from the state $c_{+}$ to $c_{-}$ in the small time interval $\Delta
t$ given by $\alpha \Delta t+o(\Delta t)$, and from the state $c_{-}$ to $c_{+}$ given by  $\beta \Delta
t+o(\Delta t)$, where $\alpha,\beta>0$. It has the correlation function
$$
\langle \xi(t)\xi(s)\rangle
=\dfrac{\alpha\beta(c_{+}-c_{-})^2}{(\alpha+\beta)^2}\exp
(-(\alpha+\beta)t).
$$
A system subject to this type of noise is described by the following equation
$$
\dfrac{dx}{dt}=F(x)+\sigma(x)\xi.
$$
The pair $(x(t),\xi(t))$ is Markovian and writing
$a(x,c_{\pm})=F(x)+c_\pm\sigma(x)$ we arrive at
$$
dx(t)= a(x(t),\xi(t))dt.
$$
The process $\xi(t)$ determines which deterministic system
$$
\dfrac{dx_+}{dt}=a(x,c_{+})\qquad\mbox{or}\qquad
\dfrac{dx_-}{dt}=a(x,c_{-})
$$
to choose. We assume that given the initial condition $x_{\pm}(0)=x$, each of these equations has a solution
$x_{+}(t)$ and  $x_{-}(t)$, respectively, defined and finite for all $t\ge 0$, and we write
$\pi_{+}^t(x)=x_{+}(t)$ and $\pi_{-}^t(x)=x_{-}(t)$. Furthermore, we assume that there is a minimal open set $X$
such that $\pi_{\pm}^t(X)\subseteq X$ for all $t>0$. Let $\B(X\times\{c_{+},c_{-}\})$ be the sigma algebra of
Borel subsets of $X\times\{c_{+},c_{-}\}$ and let $\mu$ be the product measure on $\B(X\times\{c_{+},c_{-}\})$
which, on every set of the form $B\times \{c_{\pm}\}$, is equal to the length of the set $B$. The norm of any
element $f$ of the space $L^1(X\times\{c_{+},c_{-}\},\mu)$ is equal to
$$
||f||_1=\int_X |f_{+}(x)|dx +\int_X |f_{-}(x)|dx,
$$
and we now define a semigroup of Markov operators on this space which describes the temporal evolution of the
densities of the process $(x(t),\xi(t))$.

The evolution equation for the densities
$f_{\pm}(t,x)=f(t,x,c_{\pm})$ of the process $(x(t),\xi(t))$ is of
the form
    \bea
    \frac {\partial f_+}{\partial t} &=& - \frac{\partial
    [a(x,c_+)f_+]}{\partial x} -\alpha f_{+}+ \beta f_{-}\\
    \frac {\partial f_-}{\partial t} &=& - \frac{\partial
    [a(x,c_-)f_-]}{\partial x} +\alpha f_{+}- \beta f_{-}.
    \eea
Formally writing $f=(f_{+},f_{-})^T$, we arrive at the following equation
    \be
    \frac {\partial f}{\partial t} = Af + M f
    \label{e:dnfpe}
    \ee
    where
    $$
M=\left(%
\begin{array}{cc}
  -\alpha & \beta \\
  \alpha & -\beta \\
\end{array}%
\right) \quad \mbox{and}\quad Af=\left(%
\begin{array}{c}
  - \dfrac{\partial
    [a(x,c_+)f_+]}{\partial x} \\
  - \dfrac{\partial
    [a(x,c_{-})f_{-}]}{\partial x} \\
\end{array}%
\right).
    $$
Then the operator $A$ generates a semigroup of Markov operators $T^t$ on $L^1(X\times\{c_{+},c_{-}\},\mu)$.
%
Since $M$ is a bounded operator, $A+M$ also generates a semigroup of Markov operators $P^t$, and the following
holds
$$
P^tf_0=T^tf_0+\int_0^t T^{t-s} M P^s f_0 ds
$$
for every $f_0\in L^1(X\times\{c_{+},c_{-}\},\mu)$. The semigroup $P^t$ gives the generalized solution
$f(t,\cdot)$ of Eq. \ref{e:dnfpe} with the initial condition $f(0,\cdot)=f_0$. Using the results of
\citet[Proposition 2]{pichorrudnicki00} we infer that if there is an $x_0\in X$ such that $a(x_0,c_{+})\neq
a(x_0,c_{-})$, then the semigroup $P^t$ is partially integral, and if $X$ is the minimal set such that
$\pi_{\pm}^t(X)\subseteq X$, then this semigroup can have at most one stationary density. Consequently, if a
stationary density $f_*$ exists, Theorem \ref{t:unique} implies that
    \be
    \lim_{t\to\infty}H_c(P^tf_0|f_{*})=0\label{e:mdnent1}
    \ee
for all $f_0$ with $H_c(f_0|f_{*})>-\infty$, where  the conditional entropy for $P^t$ on
$L^1(X\times\{c_{+},c_{-}\},\mu)$ is equal to
\begin{widetext}
$$
H_c(P^tf_0|f_{*})=-\int_X P^tf(x,c_{+})\log \dfrac{P^tf(x,c_{+})}{f_{*}(x,c_{+})}dx-\int_X P^tf(x,c_{-})\log
\dfrac{P^tf(x,c_{-})}{f_{*}(x,c_{-})}dx.
$$
\end{widetext} by Eq.  \ref{d:conent}.


The density of the state variable $x(t)$ is an element of the space $L^1(X,m)$, where $m$ is the Lebesque
measure on $X$. It is given by
    \be
p(t,x)=P^tf_0(x,c_{+})+P^tf_0(x,c_{-}). \label{e:denstate}
    \ee
Thus the
stationary density $f_*$ of $P^t$ gives us the stationary density of
$x(t)$
$$
p_*(x)=f_{*}(x,c_{+})+f_{*}(x,c_{-}).
$$
Since
    \be
H_c(p(t)|p_*)=-\int_X p(t,x)\log \dfrac{p(t,x)}{p_{*}(x)}\, dx\ge H_c(P^tf_0|f_*), \label{e:convercon}
    \ee
we conclude that 
    \be
\lim_{t\to\infty}H_c(p(t)|p_*)=0.\label{e:mdnent2}
    \ee

For a general one-dimensional system with dichotomous noise, one can derive a formula for the stationary
density. For the sake of clarity, we follow \cite{kitahara79} and restrict our discussion to the case of
symmetric dichotomous noise where
$$
c_{+}=-c_{-}=c \qquad \mbox{and} \qquad \alpha=\beta,
$$
and
$$
a(x,c_{+})a(x,c_{-})<0,
$$
where the zeros of $a(x,c_{+})a(x,c_{-})$ are the boundaries of $X$. Then the unique stationary density of $P^t$
is given by
\begin{widetext}
$$
f_{*}(x,c_{\pm})=K\dfrac{1}{|a(x,c_{\pm})|}\exp\left\{-\alpha \int^x \left (
\dfrac{1}{a(z,c_{+})}+\dfrac{1}{a(z,c_{-})} \right )\,dz\right\},
$$
\end{widetext} where $K$ is a normalizing constant.

As a specific example, consider the linear dichotomous flow
$$
\dfrac{dx}{dt}=-\gamma x+\xi,
$$
where $\gamma>0$.  Then $c_{\pm}=\pm c$ and $a(x,c_{\pm})=-\gamma x
\pm c$ for $x\in \realnos$. Thus
$$
\pi_{\pm}^t(x)=xe^{-\gamma t} \pm \dfrac{c}{\gamma}(1-e^{-\gamma t})
$$
and the state space is
$$
X=\left(-\dfrac{c}{\gamma}, \dfrac{c}{\gamma}\right).
$$
The stationary density $f_*$ in this case is given by
$$
f_{*}(x,\pm c)=\dfrac{K}{\mp\gamma x+c} \left(\dfrac{c^2}{\gamma^2}-x^2\right)^{\alpha/\gamma -1},
$$
where the normalizing constant is equal to
$$
K=\dfrac{\gamma}{2B\left(\frac{1}{2},\frac{\alpha}{\gamma}\right)}
\left(\dfrac{c}{\gamma}\right)^{-2\alpha/\gamma}
$$
and $B$ is the beta function. Since $a(x,+c)>0>a(x,-c)$ for all $x\in X$, Conditions \ref{e:mdnent1} and
\ref{e:mdnent2} hold, where the stationary density $p_*$ of the state variable $x(t)$ is equal to
$$
p_*(x)=\dfrac{1}{B\left(\frac{1}{2},\frac{\alpha}{\gamma}\right)}
\left(\dfrac{c}{\gamma}\right)^{1-2\alpha/\gamma}\left(\dfrac{c^2}{\gamma^2}-x^2\right)^{\alpha/\gamma -1}.
$$

\section{Discussion}\label{s:disc}

Here we have examined   the evolution of the conditional (or Kullback-Leibler
or relative) entropy to a maximum in stochastic systems. We were motivated  by
a desire to understand the role of noise in the evolution of the conditional
entropy to a maximum since  in invertible systems (e.g. measure preserving
systems of differential equations or invertible maps) the conditional entropy
is fixed at the value with which the system is prepared. However the addition
of noise can reverse this property and lead to an evolution of the conditional
entropy to a maximum value of zero. We have made  concrete calculations to see
how the entropy converges, and shown that it is  monotone and at least
exponential in several situations.

Specifically, in Section \ref{s:asre} we introduced the dynamic concept of
asymptotic stability and the notion of conditional entropy, and gave two main
results connecting the convergence of the conditional entropy with asymptotic
stability (Theorem \ref{t:entropyconv}), and the existence of unique stationary
densities with the convergence of the conditional entropy (Theorem
\ref{t:unique}). In Section \ref{s:det} we illustrated the well known fact that
asymptotic stability is a property that cannot be found in a deterministic
system of ordinary differential equations, and consequently that the
conditional entropy cannot not have time dependent behavours character for this
type of invertible dynamics. Section \ref{s:gauss} introduced  a stochastic
extension to this invertible and constant entropy situation in which a system
of ordinary equations is perturbed by Gaussian white noise (thus becoming
non-invertible).  We were able to give some general results showing that in
this stochastic case asymptotic stability holds. Then in Section \ref{ss:1d} we
considered  specific one dimensional examples, and  showed that the conditional
entropy convergence to zero is monotone and at least exponential, considering
the specific examples of an Ornstein-Uhlenbeck process in Section
\ref{sss:1dou} and a Rayleigh process in Section \ref{sss:rp}.  We went on to
look at multidimensional stochastic systems in Section \ref{ss:md}, showing
that the exponential convergence of the entropy still holds. Specific examples
of these higher dimensional situations were considered within the context of a
two dimensional Ornstein-Uhlenbeck process in Section \ref{ss:mo-up} with
specific examples of a stochastically perturbed harmonic oscillator (Section
\ref{sss:ho}) and colored noise (Section \ref{sss:cn}) as examples.  In the
specific cases of the Ornstein-Uhlenbeck and Rayleigh processes as well as the
stochastically perturbed harmonic oscillator and colored noise examples, we
have the rather surprising result that the rate of convergence of the entropy
to zero is independent of the noise amplitude $\sigma$ as long as $\sigma > 0$.
The final Section \ref{s:mdn} applied our general results to the problem of
conditional entropy convergence in the presence of dichotomous noise.


\begin{acknowledgments}
This work was supported by the Natural Sciences and Engineering Research Council (NSERC grant OGP-0036920,
Canada) and MITACS. This research was carried out while MT-K was visiting McGill University in 2004.
\end{acknowledgments}

\appendix*
\section{}
\subsection{Proof of inequality \ref{i:foot1}}

Jensen's inequality states that when $\nu$ is a normalized measure and $\phi$ is a concave function
then
$$
\phi \left (\int g(z)\nu(dz)\right )\ge \int \phi (g(z))\nu(dz).
$$
Since $\log$ is a concave function, Jensen's inequality implies
    $$
    \int f \log \frac{f}{g} \, \mu(dx)\le \log\int f\frac{f}{g} \, \mu(dx).
    $$
Now note that
    \bea
    \int \left(\frac{f}{g}-1\right)^2g \, \mu(dx)& = &\int
    \left [ \left(\frac{f}{g}\right)^2g-2\frac{f}{g}g+1\right ]
    \, \mu(dx)\nonumber\\ &=&\int\left(\frac{f}{g}\right)^2g\, \mu(dx) -1.\nonumber
    \eea
Thus
    \bea
    -H_c(f|g)&\le & \log \left [1+\int \left(\frac{f}{g}-1\right)^2g \,
    \mu(dx)\right ]\nonumber\\ &\le &\int \left(\frac{f}{g}-1\right)^2g \,
    \mu(dx).\nonumber
   \eea

\subsection{Proof of inequality \ref{e:lowbound}}
To see how easy it is to derive this estimate, first write \bea
        H_c(P^tf_1|P^tf_2) &=& -\int P^tf_1(x) \log
    \dfrac{P^tf_1(x)}{P^tf_2(x)}dx \nonumber\\&\equiv& - \int
    \varphi(P^tf_1(x),P^tf_2(x)),\nonumber
    \eea
where $\varphi(u,v) = v \log (u/v)$ is convex.  From the properties of convex functions there always exist
sequences of real numbers $\{a_n\}$ and $\{b_n\}$ such that
    $$
    \varphi(u,v) = \sup\{a_nu+b_nv: n \in \naturalnos \}.
    $$
Remembering (\ref{mo}) we can then write
\begin{widetext}
    \bea
    a_n P^tf_1(x) + b_n P^tf_2(x)
    &=&
    a_n \int k(t,x,y_1) f_1(y_1)dy_1 +
    b_n \int k(t,x,y_2) f_2(y_2)dy_2 \nonumber \\
    &=&
    \int \int \left \{ a_n k(t,x,y_1) f_1(y_1) f_2(y_2)  +
    b_n k(t,x,y_2) f_1(y_1) f_2(y_2)\right \} dy_1 dy_2 \nonumber \\
    &=&
    \int \int f_1(y_1) f_2(y_2) \left \{ a_n k(t,x,y_1) +
    b_n k(t,x,y_2) \right \} dy_1 dy_2 \nonumber \\
    &\leq &
    \int \int f_1(y_1) f_2(y_2) \varphi( k(t,x,y_1), k(t,x,y_2)) dy_1
    dy_2.\nonumber
    \eea
    \end{widetext}
Thus
\begin{widetext}
    $$
    \sup_{n \in \naturalnos} \{a_n P^tf_1(x) + b_n P^tf_2(x)\} =
    \varphi (P^tf_1(x), P^tf_2(x)) =
    \int \int f_1(y_1) f_2(y_2) \varphi( k(t,x,y_1), k(t,x,y_2)) dy_1
    dy_2,
    $$
    \end{widetext}
so
\begin{widetext}
    \bea
    H_c(P^tf_1(x)|P^tf_2(x))
    &\geq &
    - \int \int f_1(y_1) f_2(y_2) \left [ \int \varphi( k(t,x,y_1), k(t,x,y_2))dx \right ] dy_1
    dy_2 \nonumber \\
    &=&
    - \int \int f_1(y_1) f_2(y_2) H_c( k(t,x,y_1), k(t,x,y_2)) dy_1
    dy_2. \nonumber
    \eea
    \end{widetext}
\bibliography{zpf}
\end{document}